\documentclass[showpacs,preprintnumbers,prd,]{revtex4}%
\usepackage{amsfonts}
\usepackage{amsmath}
\usepackage{graphicx}
\usepackage{amssymb}
\usepackage{float}%
\providecommand{\U}[1]{\protect\rule{.1in}{.1in}}
\begin{document}
\title{Tachyon potentials from a supersymmetric FRW model}
\author{G. Garc\'ia-Jim\'enez}
\affiliation{Benem\'erita Universidad Aut\'onoma de Puebla, Facultad de Ciencias
F\'{\i}sico Matem\'aticas, P.O. Box 165, 72000 Puebla, M\'exico.}
\author{C. Ram\'{\i}rez}
\email{cramirez@fcfm.buap.mx}
\affiliation{Benem\'erita Universidad Aut\'onoma de Puebla, Facultad de Ciencias
F\'{\i}sico Matem\'aticas, P.O. Box 165, 72000 Puebla, M\'exico.}
\author{V. V\'azquez-B\'aez}
\email{manuel.vazquez@correo.buap.mx}
\affiliation{Benem\'erita Universidad Aut\'onoma de Puebla, Facultad de Ciencias
F\'{\i}sico Matem\'aticas, P.O. Box 165, 72000 Puebla, M\'exico.}

\begin{abstract}
Considering that the effective theory of closed string tachyons can have
worldsheet supersymmetry, as shown by Vafa, we study a worldline
supersymmetric action in a FRW background, whose superpotential originates
a tachyon scalar potential. There are such
potentials with spontaneously broken supersymmetry at the instability and
supersymmetric after tachyon condensation. Furthermore, given a tachyonic
potential, the superpotential can be computed by a power series ansatz and
has a free parameter which can be chosen such that complex
solutions become real.

\end{abstract}

\pacs{98.80.-k, 98.80.Qc, 04.50.Kd, 04.65.+e, 11.25.Yb, 11.27.+d}
\maketitle

\section{Introduction}

\label{intro} Many phenomena in physics are related to the transition or decay
from an unstable state, to other stable one. The corresponding evolution is
described frequently by a tachyonic potential, with a `negative' mass term,
like in the case of the Higgs potential or in the Landau-Ginzburg theory.
String theory has in its lowest mode tachyons. However the inclusion of
supersymmetry is consistent because they can be eliminated by the GSO
truncation. Nevertheless, a better knowledge of string theory requires the
understanding of the unstable configurations, whose evolution can be described
by the condensation of the tachyonic modes. The complexity of string theory
has made this study rather difficult, and it was first performed in the
somewhat simpler instance of open strings, resumed by the well known Sen
conjectures \cite{sen}. For closed strings the situation is more complicated,
in particular because it involves the structure of space-time. An interesting
fact in this case, is that closed string tachyons, which are nonsupersymmetric
in target space, can have worldsheet supersymmetry \cite{vafa}. In this sense
we address the question of supersymmetric tachyons in the simplified framework
of a FRW background, with `worldline' local supersymmetry, i.e. the time
variable is extended to the superspace of supersymmetry. 

Supersymmetric quantum
cosmology has been studied in various formulations. As usual for uniform spaces, 
it has been obtained as the ``minisuperspace"
formulation \cite{ryan} of four dimensional supergravity \cite{obregon0}, see also \cite{eath}. 
The Wheeler-deWitt equation is traced back to the `square root' of the hamiltonian constraint,  
i.e. the supersymmetric charge constraints. Additionally to these constraints, there are also the 
Lorentz transformations of the fermionic degrees of freedom \cite{obregon0}, which restrict strongly the solutions \cite{obregon}. 
An alternative supersymmetrization of these models 
has been given in \cite{superfield}, by a `worldline' one-dimensional superfield approach, 
where the time variable is extended to a (supersymmetry) superspace.
This formulation has been worked out systematically for all Bianchi models \cite{bianchi},
matter has been included as well  \cite{superfield}, see also \cite{moniz}. 
We follow an approach of this type, by means of the covariant formulation of
one-dimensional supergravity, given by the so called `new' $\Theta$ variables
\cite{wess,cupa}, which allows in a systematic and straightforward way to write supergravity
invariant actions. The fermionic sector of the superfield approach has less components than the one of the reduction of four dimensional supergravity, as it is one dimensional, which is some sense compensates the absence of Lorentz constraints. In fact, in both approaches the wave function of the universe seems to have only two independent components \cite{tkachwf,obregon}. However, the complexity of four, or higher dimensional supergravity, has made interesting the study of the worldline superfield approach, which keeps the essence of supersymmetry in a relative simple way.

The action we are considering contains two real scalars, one of them the
dilaton, and the other one has a tachyonic potential $V(T)$ \cite{zwiebach},
coupled to FRW supergravity. We formulate one-dimensional $N=2$ supespace
supergravity following \cite{cupa} and the superfield form of the action is
taken from \cite{superfield}. The final action is obtained after
a rescaling and eliminating the auxiliary fields. For completeness we give
also the hamiltonian formulation which closes consistently without further
complications. As usual in supergravity, the superpotential is related to the
scalar potential by a differential equation which is not positive definite. In
order to solve this equation, we consider the case $k=0$ and make an ansatz of
separation of variables. We look for superpotentials corresponding to
tachyonic potentials $V(T)$ and in particular such that both supersymmetries
are spontaneously broken at the maximum, and after condensation they are
restored. We look also for a general solution of this
differential equation by a power series ansatz. With this ansatz, depending on
the potential, the superpotential can be complex, with complex values for
quantities like the mass. However there is a parameter which can be chosen in
such a way that the solutions are real. In the second section of this work we
give our start point bosonic action, in the third section we give the
one-dimensional superspace supergravity formulation, in the fifth section we
give the tachyon action, in the sixth section the hamiltonian is formulated
and in the next section we study the solutions for the superpotential, in the
last section we sketch conclusions. There are two appendices, the first one on
the `new' $\Theta$-variables formulation and in the second one there are
details of the power series solutions.


\section{Closed String Tachyon Effective Action}

The closed string tachyon effective action in the bosonic sector is given
according to \cite{zwiebach} as%
\begin{equation}
S=\frac{1}{2\kappa_{D}^{2}}\int\sqrt{-g}e^{-2\phi}\left[  R+4\left(
\partial\phi\right)  ^{2}-\left(  \partial T\right)  ^{2}-2V(T)\right]
d^{D}x, \label{action}%
\end{equation}
where $T$ is the closed string tachyon field, $V(T)$ is the tachyon potential
and $\phi$ is the dilaton field. This action can be written in the Einstein
frame by means of $g_{\mu\nu}^{string}=e^{\phi}g_{\mu\nu}^{Einstein} $, which
is more suitable for our cosmological approach. For a four dimensional FRW
metric and in the Einstein frame, the Lagrangian takes the form%
\begin{equation}
S=\int\left[  -\frac{3\dot{a}^{2}a}{\kappa^{2}N}+\frac{3Nka}{\kappa^{2}}%
+\frac{a^{3}}{\kappa^{2}N}\dot{\phi}^{2}+\frac{a^{3}}{2\kappa^{2}N}\dot{T}%
^{2}-\frac{a^{3}Ne^{2\phi}V\left(  T\right)  }{\kappa^{2}}\right]  dt,
\label{actioncosmo}%
\end{equation}
where $N$ is the lapse function and $a$ is the scale factor. This Lagrangian
is invariant (up to a total derivative) under time reparametrizations of the
form $t\rightarrow f(t)$. This invariance under time reparametrization is
extended to supersymmetry supersymmetry by the introduction of a Grassmann
superspace associated to the bosonic time coordinate $t$ (see Tkach et al. in
\cite{superfield}).

As usual, the Hamiltonian of the bosonic theory, has the form $H=NH_{0}$ where
$N$ is the lapse function. Then, the associated equation of motion\ $\partial
H/\partial N=0$ implies the first class constraint%
\begin{equation}
H_{0}=-\frac{\kappa^{2}}{12a}\pi_{a}^{2}-\frac{3ka}{\kappa^{2}}+\frac
{\kappa^{2}}{4a^{3}}\pi_{\phi}^{2}+\frac{\kappa^{2}}{2a^{3}}\pi_{T}^{2}%
+\frac{a^{3}e^{2\phi}V(T)}{\kappa^{2}}=0, \label{bosonichamiltonian}%
\end{equation}
where $\pi_{i}$ are the canonical momenta of the coordinates $i=a,T,\phi$.


\section{Superspace supergravity}

Superspace is the natural framework for a geometrical formulation of
supersymmetry and supergravity \cite{salam}. It extends spacetime by
anticommuting Grassmann variables, $x^{m}\rightarrow(x^{m},\theta^{\mu})$. The
field content of the superfields is given by the Grassmann power expansion in
the anticommuting variables $\phi(z)=\sum_{n}1/n!\, \theta^{\mu_{1}}%
\cdots\theta^{\mu_{n}}\phi_{\mu_{1}\cdots\mu_{n}}(x)$. Supergravity is
invariant under local supersymmetry transformations $\xi^{m} \rightarrow
\xi^{m}(x)$, $\xi^{\mu}\rightarrow\xi^{\mu}(x)$, where $\xi^{m}$ are spacetime
translation and $\xi^{\mu}$ supersymmetry transformation parameters. It can be
generalized to superspace diffeomorphisms \cite{wess}, $z^{M} \equiv
(x^{m},\theta^{\mu}) \rightarrow{z^{\prime}}^{M}= z^{M}+ \xi^{M}(z)$. This
generalization actually amounts to introduce additional `superspace' gauge
degrees of freedom corresponding to the $\theta$-components of the Grassmann
expansion of $\xi^{M}(x,\theta)$. In order to formulate such a theory, the
vierbein and spin connection are generalized to superspace tensors, the
vielbein ${E_{M}}^{A}(z)$, and the superconnection ${\phi_{MA}}^{B}$, where
$A=(a, \alpha)$ are local Lorentz indices and $M=(m,\mu) $ are superspace
world indices \cite{wess}. If $V_{A}$ is a Lorentz supervector, its covariant
derivatives are $\mathcal{D}_{A} V_{B} ={E_{A}}^{M}(\partial_{M} V_{A}-
{\phi_{M B}}^{C} V_{C})$, and satisfy the graded (anti)commutators
$[\mathcal{D}_{A},\mathcal{D}_{B}]_{\pm}V_{C}=-{T_{AB}}^{D}\mathcal{D}%
_{D}V_{C}-{R_{ABC}}^{D}V_{D}$, where ${E_{A}}^{M}(z)$ is the inverse vielbein
and the torsion and curvature tensors satisfy the graded Bianchi identitites.
For the construction of the lagrangian, in \cite{superfield} a somewhat adhoc
superspace formulation has been proposed. Here we will use the `new'
superspace formulation, see Apendix \ref{apsugra}, which follows from general
superspace covariance, by a consistent elimination of the superspace gauge
degrees of freedom \cite{wess,new,cupa}, without requiring a gauge fixing.
This parametrization corresponds to a field redefinition of the superfield
components $\phi_{\mu_{1}\cdots\mu_{n}}(x)\rightarrow\left.  \mathcal{D}%
_{[\alpha_{1}}\cdots\mathcal{D}_{\alpha_{n}]} \phi(z)\right|  _{\theta=0}$ and
the superfields are given by $\Phi(x,\Theta)=\left.  e^{\Theta^{\alpha
}\mathcal{D} _{\alpha}} \phi(z)\right|  _{\theta=0}$, where $\Theta$ are
anticommuting Lorentz spinor variables. Full manifest covariance can be
mantained by keeping $\Phi(z,\Theta)= e^{\Theta^{\alpha}\mathcal{D} _{\alpha}}
\phi(z)$ and setting $\theta=0$ at the end of the computations. In this
formulation, the supergravity multiplet contains the vierbein, the spin
conection, and certain components of the curvature and torsion tensors,
constrained by the Bianchi identities. In Appendix \ref{apsugra} this
formulation is reviewed, following \cite{cupa}. Superfields transform by field
dependent transformations (\ref{trafos}) and covariant derivatives can be
defined consistently (\ref{covder}) in such a way that there is a vielbein
whose superdeterminant is an invariant density (\ref{trafoe}) and allows to
construct invariant supergravity actions. We use these results for a
superspace formulation of one-dimensional supergravity. To set conventions, we
first observe that simple one-dimensional supersymmetry has the grassmanian
variables $\theta$ and $\bar{\theta}=\theta^{\dagger}$. The integration
properties for these variables are $\int d\theta=0$, $\int\theta d\theta=1$.
Generic superfields are real and are given by $\phi(t,\theta,\bar{\theta
})=A(t)+\theta\psi(t)-\bar{\theta}\bar{\psi}(t)+\theta\bar{\theta}B(t)$, where
$A(t)$ and $B(t)$ are real and $\psi^{\dagger}=\bar{\psi}$. The representation
of supersymmetric charges is $Q=\frac{d}{d\theta}-i\bar{\theta}\frac{d}{dt}$
and $\bar{Q}=-\frac{d}{d\bar{\theta}}+i\theta\frac{d}{dt}$, which satisfy
$\{Q,\bar{Q}\}=2i\frac{d}{dt}$, the covariant derivatives are $D=\frac
{d}{d\theta}+i\bar{\theta}\frac{d}{dt}$ and $\bar{D}=-\frac{d}{d\bar{\theta}%
}-i\theta\frac{d}{dt}$. There are also chiral superfields which are complex
and satisfy $\bar{D}\phi=0$, and which in the chiral base have the expansion
$\phi(t,\theta)=A(t)+\theta\psi(t)$.

Supergravity in one dimension, as well as gravity, is trivial and there is no
curvature tensor, and in a minimal version, the torsion tensor is the same as
in flat superspace, i.e. its only non-vanishing component is ${T_{\theta
\bar\theta}}^{t}=2i$, consistently with the Bianchi identities, as can be
easily verified by dimensional reduction from minimal supergravity in
four dimensions \cite{wess}, or by direct computation in one dimension. For
the one-dimensional `new' superspace formulation, unlike the appendix
\ref{apsugra}, we will denote $z^{M}=(t,\Theta,\bar{\Theta})$ and for
simplicity we will omit the tildes. Thus superfields $\Phi(z)=A(t)+\Theta
\chi(t)-\bar{\Theta}\bar{\chi}(t)+\Theta\bar{\Theta}B(t)$ transform as
\begin{equation}
\delta_{\xi}\Phi(z)={\eta_{\xi}}^{M}(z)\partial_{M}\Phi(z)\equiv\xi
^{M}(x)\left[  \eta_{M}^{\ t}(z)\frac{\partial\Phi(z)}{\partial t}+\eta
_{M}^{\ \theta}(z)\frac{\partial\Phi(z)}{\partial\Theta}+\eta_{M}%
^{\ \bar\theta}(z)\frac{\partial\Phi(z)}{\partial\bar\Theta}\right]  ,
\label{trafophi}%
\end{equation}
Thus, from (\ref{eta}) it turns out that
\begin{align}
{\eta_{\xi}}^{t}  &  =-\xi-ie^{-1}\left(  \Theta\bar{\zeta}+\bar{\Theta}%
\zeta\right)  +\frac{1}{2}e^{-2}\Theta\bar{\Theta}\left(  \bar{\zeta}%
\psi-\zeta\bar{\psi}\right)  ,\\
{\eta_{\xi}}^{\theta}  &  =-\zeta+\frac{i}{2}e^{-1}\left(  \Theta\bar{\zeta
}+\bar{\Theta}\zeta\right)  \psi-\frac{1}{4}e^{-2}\Theta\bar{\Theta}\zeta
\psi\bar{\psi},
\end{align}
where the supersymmetry parameters are $\xi^{M}(x)=(\xi^{t},\xi^{\theta}%
,\xi^{\bar{\theta}})\equiv(\xi,\zeta,\bar{\zeta})$ and ${\eta_{\xi}}%
^{\bar{\theta}}=({\eta_{\xi}}^{\theta})^{\dagger}$. The resulting component
transformations are
\begin{align}
\delta_{\xi}A  &  =-\xi\dot{A}-\zeta\chi+\bar{\zeta}\bar{\chi},\\
\delta_{\xi}\chi &  =-\xi\dot{\chi}-ie^{-1}\bar{\zeta}\dot{A}+\frac{i}%
{2}e^{-1}\bar{\zeta}(\psi\chi-\bar{\psi}\bar{\chi})+\bar{\zeta}B,
\label{trafochi}\\
\delta_{\xi}\bar\chi &  =-\xi\dot{\bar\chi}+ie^{-1}{\zeta}\dot{A}-\frac{i}%
{2}e^{-1}{\zeta}(\psi\chi-\bar{\psi}\bar{\chi})+{\zeta}B,
\label{trafochibar}\\
\delta_{\xi}B  &  =-\xi\dot{B}-ie^{-1}(\zeta\dot{\chi}+\bar{\zeta}\dot
{\bar{\chi}})+\frac{1}{2}e^{-2}\zeta\bar{\psi}(\dot{A}+iB)-\frac{1}{2}%
e^{-2}\bar{\zeta}\psi(\dot{A}-iB)-\frac{1}{12}e^{-2}(\zeta\chi-\bar{\zeta}%
\bar{\chi})\psi\bar{\psi},
\end{align}
as well as the one of $\bar{\chi}$, obtained from (\ref{trafochi}) by complex
conjugation. Further, the vielbein corresponding to transformations
(\ref{trafophi}), which transforms as $\delta_{\xi}{\nabla_{M}}^{A}%
=\partial_{M}\eta_{\xi}^{N}{\nabla_{N}}^{A}+\eta_{\xi}^{N}\partial_{N}%
{\nabla_{M}}^{A}$, can be obtained from (\ref{nabla})
\begin{equation}
{\nabla_{M}}^{A}=\left(
\begin{array}
[c]{ccc}%
e+i(\Theta\bar{\Psi}+\bar{\Theta}\psi) & \frac{1}{2}\psi & \frac{1}{2}%
\bar{\psi}\\
-i\bar{\Theta} & 0 & -1\\
i\Theta & 1 & 0
\end{array}
\right) , \label{vielbein}%
\end{equation}
and from its transformation law we get
\begin{align}
\delta_{\xi}e  &  =-\frac{d}{dt}(\xi e)+i(\zeta\bar{\psi}+\bar{\zeta}\psi),\\
\delta_{\xi}\psi &  =-2\frac{d}{dt}\left(  \zeta+\frac{1}{2}\xi\psi\right)  ,
\end{align}
which can be verified to be consistent with the usual vielbein transformations
\cite{wess}. The invariant density is obtained as usual from the
superdeterminant $\mathcal{E}=\mathrm{Sdet}({\nabla_{M}}^{A})$, and transforms
as $\delta_{\xi}\mathcal{E}=(-1)^{m}\partial_{M}(\xi^{M}\mathcal{E})$
\begin{equation}
\mathcal{E}=-e-\frac{i}{2}(\Theta\bar{\Psi}+\bar{\Theta}\psi),
\end{equation}
The inverse vielbein can be computed from (\ref{vielbein}), and from it we get
the covariant derivatives which will be needed for the lagrangian of the next
section
\begin{align}
\nabla_{\theta}\Phi &  =\chi+ie^{-1}\bar{\Theta}\left[  \dot{A}-\frac{1}%
{2}(\psi\chi-\bar{\psi}\bar{\chi})-ieB\right]  +e^{-1}\Theta\bar{\Theta
}\left(  -i\dot{\chi}-\frac{1}{2}e^{-1}\bar{\psi}\dot{A}-\frac{1}{4}e^{-1}%
\psi\bar{\psi}\chi+\frac{i}{2}\bar{\psi}B\right)  ,\\
\nabla_{\bar{\theta}}\Phi &  =\bar{\chi}-ie^{-1}\Theta\left[  \dot{A}-\frac
{1}{2}(\psi\chi-\bar{\psi}\bar{\chi})+ieB\right]  +e^{-1}\Theta\bar{\Theta
}\left(  i\dot{\bar{\chi}}-\frac{1}{2}e^{-1}\psi\dot{A}-\frac{1}{4}e^{-1}%
\psi\bar{\psi}\bar{\chi}-\frac{i}{2}\psi B\right)  .
\end{align}


\section{Susy Closed Tachyon Model}

The supersymmetric cosmological model is obtained upon an extension of the
time coordinate into a supermultiplet $t\rightarrow(t,\Theta,\bar{\Theta})$.
Due to this time supersymmetric generalization, also the fields of the theory
are generalized as superfields, the expansion is given by%
\begin{equation}%
\begin{array}
[c]{c}%
\mathcal{A}\left(  t,\Theta,\bar{\Theta}\right)  =a\left(  t\right)
+i\Theta\bar{\lambda}\left(  t\right)  +i\bar{\Theta}\lambda\left(  t\right)
+B\left(  t\right)  \Theta\bar{\Theta},\\
\mathcal{T}\left(  t,\Theta,\bar{\Theta}\right)  =T\left(  t\right)
+i\Theta\bar{\eta}\left(  t\right)  +i\bar{\Theta}\eta(t)+G\left(  t\right)
\Theta\bar{\Theta},\\
\Phi\left(  t,\Theta,\bar{\Theta}\right)  =\phi\left(  t\right)  +i\Theta
\bar{\chi}\left(  t\right)  +i\bar{\Theta}\chi\left(  t\right)  +F\left(
t\right)  \Theta\bar{\Theta},
\end{array}
\label{superfields}%
\end{equation}
where, $\mathcal{A}$, $\mathcal{T}$\ and $\Phi$ are the superfields of $a$,
$T$ and $\phi$.

The supersymmetric generalization of the action is given by%
\begin{equation}
S=S_{Rsusy}+S_{Msusy}, \label{susyaction}%
\end{equation}
where, $S_{Rsusy}$ is the cosmological supersymmetric generalization of the
free FRW model
\begin{equation}
S_{Rsusy}=\int\left(  \frac{3\mathcal{E}}{\kappa^{2}}\mathcal{A}\nabla
_{\bar{\theta}}\mathcal{A}\nabla_{\theta}\mathcal{A}-\frac{3\sqrt{k}}%
{\kappa^{2}}\mathcal{EA}^{2}\right)  d\Theta d\bar{\Theta}dt,
\label{susyraction}%
\end{equation}
and the supersymmetric matter term is%
\begin{align}
S_{Msusy}  &  =\frac{1}{\kappa^{2}}\int\left[  -\mathcal{E}\mathcal{A}%
^{3}\nabla_{\bar{\Theta}}\Phi\nabla_{\Theta}\Phi-\frac{1}{2}\mathcal{EA}%
^{3}\nabla_{\bar{\Theta}}\mathcal{T}\nabla_{\Theta}\mathcal{T}\right.
\label{susymaction}\\
&  \left.  +\mathcal{EA}^{3}W\left(  \Phi,\mathcal{T}\right)  \right]  d\Theta
d\bar{\Theta}dt,\nonumber
\end{align}
where $W\left(  \Phi,\mathcal{T}\right)  $ is the superpotential. The
superpotential expansion can be written as $W\left(  \Phi,\mathcal{T}\right)
=W(\phi,T)+\frac{\partial W}{\partial\phi}(\Phi-\phi)+\frac{\partial
W}{\partial T}(\mathcal{T}-T)+\frac{1}{2}\frac{\partial^{2}W}{\partial T^{2}%
}(\mathcal{T}-T)^{2}+\frac{1}{2}\frac{\partial^{2}W}{\partial\phi^{2}}%
(\Phi-\phi)^{2}+\frac{\partial^{2}W}{\partial T\partial\phi}(\mathcal{T}%
-T)(\Phi-\phi)$. This expansion is finite because the terms $(\mathcal{T}-T)$
and $(\Phi-\phi)$ are purely grassmannian. Upon integration over the Grassmann
parameters, we find the supersymmetric cosmological Lagrangian in the gravity
sector%
\begin{align*}
L_{FRWsusy}  &  =-\frac{3a\dot{a}^{2}}{e\kappa^{2}}+\frac{3ia}{\kappa^{2}%
}\left(  \lambda\dot{\bar{\lambda}}-\dot{\lambda}\bar{\lambda}\right)
+\frac{6e\sqrt{k}\lambda\bar{\lambda}}{\kappa^{2}}+\frac{3a\dot{a}}%
{e\kappa^{2}}\left(  \psi\lambda-\bar{\psi}\bar{\lambda}\right)
+\frac{3ia\sqrt{k}}{\kappa^{2}}\left(  \psi\lambda+\bar{\psi}\bar{\lambda
}\right) \\
&  -\frac{3aB^{2}e}{\kappa^{2}}+\frac{6aBe\sqrt{k}}{\kappa^{2}}-\frac
{3Be\lambda\bar{\lambda}}{\kappa^{2}}-\frac{3a}{2e\kappa^{2}}\lambda
\bar{\lambda}\psi\bar{\psi},
\end{align*}
the dilaton sector of the matter Lagrangian is given by%
\begin{align*}
L_{\phi susy}  &  =\frac{a^{3}\dot{\phi}^{2}}{e\kappa^{2}}+\frac{a^{3}%
\dot{\phi}}{e\kappa^{2}}\left(  \bar{\psi}\bar{\chi}-\psi\chi\right)
+\frac{ia^{3}}{\kappa^{2}}\left(  \dot{\chi}\bar{\chi}-\chi\dot{\bar{\chi}%
}\right)  +\frac{3ia^{2}\dot{\phi}}{\kappa^{2}}\left(  \lambda\bar{\chi}%
+\bar{\lambda}\chi\right) \\
&  +\frac{a^{3}eF^{2}}{\kappa^{2}}+\frac{3a^{2}eF}{\kappa^{2}}\left(
\lambda\bar{\chi}-\bar{\lambda}\chi\right)  -\frac{3a^{2}Be\chi\bar{\chi}%
}{\kappa^{2}}+\frac{a^{3}\chi\bar{\chi}\psi\bar{\psi}}{2e\kappa^{2}}%
-\frac{6ae\lambda\bar{\lambda}\chi\bar{\chi}}{\kappa^{2}},
\end{align*}
while for the tachyonic sector we find%
\begin{align*}
L_{Tsusy}  &  =\frac{a^{3}\dot{T}^{2}}{2e\kappa^{2}}+\frac{a^{3}\dot{T}%
}{2e\kappa^{2}}\left(  \bar{\psi}\bar{\eta}-\psi\eta\right)  +\frac{ia^{3}%
}{2\kappa^{2}}\left(  \dot{\eta}\bar{\eta}-\eta\dot{\bar{\eta}}\right)
+\frac{3ia^{2}\dot{T}}{2\kappa^{2}}\left(  \lambda\bar{\eta}+\bar{\lambda}%
\eta\right) \\
&  +\frac{a^{3}eG^{2}}{2\kappa^{2}}+\frac{3a^{2}eG}{2\kappa^{2}}\left(
\lambda\bar{\eta}-\bar{\lambda}\eta\right)  -\frac{3a^{2}Be\eta\bar{\eta}%
}{2\kappa^{2}}+\frac{a^{3}\eta\bar{\eta}\psi\bar{\psi}}{4e\kappa^{2}}%
-\frac{3ae\lambda\bar{\lambda}\eta\bar{\eta}}{\kappa^{2}},
\end{align*}
and the superpotential term%
\begin{align*}
L_{W}  &  =-\frac{6aeW\lambda\bar{\lambda}}{\kappa^{2}}-\frac{3ia^{2}%
W}{2\kappa^{2}}\left(  \bar{\psi}\bar{\lambda}+\psi\lambda\right)
-\frac{3a^{2}BeW}{\kappa^{2}}+\frac{3a^{2}eW_{T}}{\kappa^{2}}\left(
\bar{\lambda}\eta-\lambda\bar{\eta}\right) \\
&  -\frac{ia^{3}W_{T}}{2\kappa^{2}}\left(  \bar{\psi}\bar{\eta}+\psi
\eta\right)  -\frac{a^{3}eGW_{T}}{\kappa^{2}}+\frac{3a^{2}eW_{\phi}}{\kappa
^{2}}\left(  \bar{\lambda}\chi-\lambda\bar{\chi}\right)  -\frac{ia^{3}W_{\phi
}}{2\kappa^{2}}\left(  \bar{\psi}\bar{\chi}+\psi\chi\right) \\
&  -\frac{a^{3}eFW_{\phi}}{\kappa^{2}}-\frac{a^{3}e\chi\bar{\chi}W_{\phi\phi}%
}{\kappa^{2}}+\frac{a^{3}eW_{T\phi}}{\kappa^{2}}\left(  \bar{\chi}\eta
-\chi\bar{\eta}\right)  -\frac{a^{3}e\eta\bar{\eta}W_{TT}}{\kappa^{2}}.
\end{align*}
Thus the total Lagrangian is
\begin{align*}
L  &  =-\frac{3a\dot{a}^{2}}{e\kappa^{2}}+\frac{3a\dot{a}}{e\kappa^{2}}\left(
\psi\lambda-\bar{\psi}\bar{\lambda}\right)  +\frac{a^{3}\dot{T}^{2}}%
{2e\kappa^{2}}-\frac{a^{3}\dot{T}}{2e\kappa^{2}}\left(  \psi\eta-\bar{\psi
}\bar{\eta}\right)  +\frac{3ia^{2}\dot{T}}{2\kappa^{2}}\left(  \lambda
\bar{\eta}+\bar{\lambda}\eta\right)  +\frac{a^{3}\dot{\phi}^{2}}{e\kappa^{2}%
}\\
&  -\frac{a^{3}\dot{\phi}}{e\kappa^{2}}\left(  \psi\chi-\bar{\psi}\bar{\chi
}\right)  +\frac{3ia^{2}\dot{\phi}}{\kappa^{2}}\left(  \lambda\bar{\chi}%
+\bar{\lambda}\chi\right)  +\frac{3ia}{\kappa^{2}}\left(  \lambda\dot
{\bar{\lambda}}+\bar{\lambda}\dot{\lambda}\right)  -\frac{ia^{3}}{2\kappa^{2}%
}\left(  \eta\dot{\bar{\eta}}+\bar{\eta}\dot{\eta}\right)  -\frac{ia^{3}}%
{\kappa^{2}}\left(  \chi\dot{\bar{\chi}}+\bar{\chi}\dot{\chi}\right) \\
&  +\frac{6e\sqrt{k}\lambda\bar{\lambda}}{\kappa^{2}}+\frac{3ia\sqrt{k}%
}{\kappa^{2}}\left(  \psi\lambda+\bar{\psi}\bar{\lambda}\right)
-\frac{6aeW\lambda\bar{\lambda}}{\kappa^{2}}-\frac{3ia^{2}W}{2\kappa^{2}%
}\left(  \psi\lambda+\bar{\psi}\bar{\lambda}\right)  -\frac{ia^{3}W_{T}%
}{2\kappa^{2}}\left(  \psi\eta+\bar{\psi}\bar{\eta}\right) \\
&  +\frac{3a^{2}eW_{T}}{\kappa^{2}}\left(  \bar{\lambda}\eta-\lambda\bar{\eta
}\right)  -\frac{ia^{3}W_{\phi}}{2\kappa^{2}}\left(  \psi\chi+\bar{\psi}%
\bar{\chi}\right)  +\frac{3a^{2}eW_{\phi}}{\kappa^{2}}\left(  \bar{\lambda
}\chi-\lambda\bar{\chi}\right)  -\frac{a^{3}eW_{TT}}{\kappa^{2}}\eta\bar{\eta
}\\
&  +\frac{a^{3}eW_{T\phi}}{\kappa^{2}}\left(  \bar{\chi}\eta-\chi\bar{\eta
}\right)  -\frac{a^{3}eW_{\phi\phi}}{\kappa^{2}}\chi\bar{\chi}-\frac
{3a\psi\bar{\psi}\lambda\bar{\lambda}}{2e\kappa^{2}}+\frac{a^{3}\psi\bar{\psi
}\eta\bar{\eta}}{4e\kappa^{2}}+\frac{a^{3}\chi\bar{\chi}\psi\bar{\psi}%
}{2e\kappa^{2}}-\frac{3ae\lambda\bar{\lambda}\eta\bar{\eta}}{\kappa^{2}}\\
&  -\frac{6ae\lambda\bar{\lambda}\chi\bar{\chi}}{\kappa^{2}}-\frac{3aB^{2}%
e}{\kappa^{2}}+\frac{6aBe\sqrt{k}}{\kappa^{2}}-\frac{3Be\lambda\bar{\lambda}%
}{\kappa^{2}}-\frac{3a^{2}Be\eta\bar{\eta}}{2\kappa^{2}}-\frac{3a^{2}%
Be\chi\bar{\chi}}{\kappa^{2}}-\frac{3a^{2}BeW}{\kappa^{2}}\\
&  +\frac{a^{3}eG^{2}}{2\kappa^{2}}+\frac{3a^{2}eG}{2\kappa^{2}}\left(
\lambda\bar{\eta}-\bar{\lambda}\eta\right)  -\frac{a^{3}eGW_{T}}{\kappa^{2}%
}+\frac{a^{3}eF^{2}}{\kappa^{2}}+\frac{3a^{2}eF}{\kappa^{2}}\left(
\lambda\bar{\chi}-\bar{\lambda}\chi\right)  -\frac{a^{3}eFW_{\phi}}{\kappa
^{2}},
\end{align*}
where the subscripts in $W$ denote partial differentiation with respect to
$\phi$ and $T$ respectively. When we perform the variation of the lagrangian
with respect to the fields $B$, $F$ and $G$, as usual the following algebraic
constraints are obtained,%
\begin{eqnarray}
B&=&\sqrt{k}-\frac{aW}{2}-\frac{1}{2a}\lambda\bar{\lambda}-\frac{1}{4}a\eta\bar{\eta}-\frac
{1}{2}a\chi\bar{\chi},\nonumber\\
G&=&W_{T}-\frac{3}{2a}\left(  \lambda\bar{\eta}-\bar{\lambda}\eta\right)  ,\label{eqaux}\\
F&=&\frac{W_{\phi}}{2}-\frac{3}{2a}\left(  \lambda\bar{\chi}-\bar{\lambda}%
\chi\right)  ,\nonumber
\end{eqnarray}
that is $B$, $F$ and $G$ play the role of auxiliary fields, and they can be
solved and eliminated from the lagrangian. When we solve for the auxiliary
fields and make the further rescalings $\lambda\rightarrow\kappa
a^{-1/2}\lambda$, $\bar{\lambda}\rightarrow\kappa a^{-1/2}\bar{\lambda}$,
$\eta\rightarrow\kappa a^{-3/2}\eta$, $\bar{\eta}\rightarrow\kappa
a^{-3/2}\bar{\eta}$, $\chi\rightarrow\kappa a^{-3/2}\chi$, $\bar{\chi
}\rightarrow\kappa a^{-3/2}\bar{\chi}$, we find the Lagrangian%
\begin{align*}
L  &  =-\frac{3a\dot{a}^{2}}{e\kappa^{2}}+\frac{3\sqrt{a}\dot{a}}{e\kappa
}\left(  \psi\lambda-\bar{\psi}\bar{\lambda}\right)  +\frac{3eka}{\kappa^{2}%
}+\frac{\dot{T}^{2}a^{3}}{2e\kappa^{2}}-\frac{\sqrt{a^{3}}\dot{T}}{2e\kappa
}\left(  \psi\eta-\bar{\psi}\bar{\eta}\right)  +\frac{3i\dot{T}}{2}\left(
\lambda\bar{\eta}+\bar{\lambda}\eta\right) \\
&  +\frac{\dot{\phi}^{2}a^{3}}{e\kappa^{2}}-\frac{\sqrt{a^{3}}\dot{\phi}%
}{e\kappa}\left(  \psi\chi-\bar{\psi}\bar{\chi}\right)  +3i\dot{\phi}\left(
\lambda\bar{\chi}+\bar{\lambda}\chi\right)  +3i\left(  \lambda\dot
{\bar{\lambda}}+\bar{\lambda}\dot{\lambda}\right)  -\frac{i}{2}\left(
\eta\dot{\bar{\eta}}+\bar{\eta}\dot{\eta}\right) \\
&  -i\left(  \chi\dot{\bar{\chi}}+\bar{\chi}\dot{\chi}\right)  +\frac
{3e\sqrt{k}\lambda\bar{\lambda}}{a}-\frac{3e\sqrt{k}\eta\bar{\eta}}{2a}%
-\frac{3e\sqrt{k}\chi\bar{\chi}}{a}+\frac{3i\sqrt{ak}}{\kappa}\left(
\psi\lambda+\bar{\psi}\bar{\lambda}\right) \\
&  +\frac{3eW^{2}a^{3}}{4\kappa^{2}}-\frac{3e\sqrt{k}a^{2}W}{\kappa^{2}}%
-\frac{eW_{T}^{2}a^{3}}{2\kappa^{2}}-\frac{eW_{\phi}^{2}a^{3}}{4\kappa^{2}%
}-\frac{9}{2}eW\lambda\bar{\lambda}+\frac{3}{4}eW\eta\bar{\eta}+\frac{3}%
{2}eW\chi\bar{\chi}\\
&  -\frac{3ia^{3/2}W}{2\kappa}\left(  \psi\lambda+\bar{\psi}\bar{\lambda
}\right)  -\frac{i\sqrt{a^{3}}W_{T}}{2\kappa}\left(  \psi\eta+\bar{\psi}%
\bar{\eta}\right)  +\frac{3eW_{T}}{2}\left(  \bar{\lambda}\eta-\lambda\bar
{\eta}\right)  -\frac{i\sqrt{a^{3}}W_{\phi}}{2\kappa}\left(  \psi\chi
+\bar{\psi}\bar{\chi}\right) \\
&  +\frac{3eW_{\phi}}{2}\left(  \bar{\lambda}\chi-\lambda\bar{\chi}\right)
-eW_{TT}\eta\bar{\eta}+eW_{T\phi}\left(  \bar{\chi}\eta-\chi\bar{\eta}\right)
-eW_{\phi\phi}\chi\bar{\chi}+\frac{3e\kappa^{2}}{4a^{3}}\eta\bar{\eta}\chi
\bar{\chi}-\frac{3}{2e}\psi\bar{\psi}\lambda\bar{\lambda}\\
&  +\frac{1}{2e}\psi\bar{\psi}\chi\bar{\chi}+\frac{1}{4e}\psi\bar{\psi}%
\eta\bar{\eta}.
\end{align*}

Substituting the equations of motion of the auxiliary fields (\ref{eqaux}) into the supersymetry
transformation of the fermions $\lambda$, $\eta$ and $\chi$ from (\ref{trafochi}), we get $\delta_\zeta\lambda=\bar\zeta\left(\sqrt{k}-aW/2\right)+\cdots$, $\delta\eta=\bar\zeta W_T+\cdots$ and $\delta\chi=\bar\zeta W_\phi+\cdots$. Therefore, if any the fields on the r.h.s. of these equations has nonvanishing v.e.v., the corresponding fermion is a goldstino and supersymmetry is broken. In the case of $\lambda$, the breaking can be due to the cosmological constant or to a nonvanishing $W$. In fact, if the superpotential has the form $W\sim e^\phi f(T)$, as in the examples in the next section, then $W_\phi=W$, i.e. $\chi$ contributes to the goldstino if $W\neq 0$.

\section{Hamiltonian analysis}

The canonical momenta are%
\begin{align*}
\pi_{a}  &  =-\frac{6a\dot{a}}{e\kappa^{2}}-\frac{3\sqrt{a}\bar{\psi}%
\bar{\lambda}}{e\kappa}+\frac{3\sqrt{a}\psi\lambda}{e\kappa},\\
\pi_{T}  &  =\frac{a^{3}\dot{T}}{e\kappa^{2}}+\frac{\sqrt{a^{3}}\bar{\psi}%
\bar{\eta}}{2e\kappa}-\frac{\sqrt{a^{3}}\psi\eta}{2e\kappa}+\frac
{3a^{3/2}i\lambda\bar{\eta}}{2\sqrt{a^{3}}}+\frac{3a^{3/2}i\bar{\lambda}\eta
}{2\sqrt{a^{3}}},\\
\pi_{\phi}  &  =\frac{2a^{3}\dot{\phi}}{e\kappa^{2}}+\frac{\sqrt{a^{3}}%
\bar{\psi}\bar{\chi}}{e\kappa}-\frac{\sqrt{a^{3}}\psi\chi}{e\kappa}%
+\frac{3a^{3/2}i\lambda\bar{\chi}}{\sqrt{a^{3}}}+\frac{3a^{3/2}i\bar{\lambda
}\chi}{\sqrt{a^{3}}},\\
\pi_{\lambda}  &  =-3i\bar{\lambda},\text{\ \ }\pi_{\bar{\lambda}}%
=-3i\lambda,\\
\pi_{\eta}  &  =\frac{i}{2}\bar{\eta},\text{\ \ }\pi_{\bar{\eta}}=\frac{i}%
{2}\eta,\\
\pi_{\chi}  &  =i\bar{\chi},\text{\ \ }\pi_{\bar{\chi}}=i\chi.
\end{align*}

As usual, we can see the appearence of the fermionic constraints%
\begin{align}
\Omega_{\lambda}  &  =\pi_{\lambda}+3i\bar{\lambda},\text{ \ }\Omega
_{\bar{\lambda}}=\pi_{\bar{\lambda}}+3i\lambda,\nonumber\\
\Omega_{\eta}  &  =\pi_{\eta}-\frac{i}{2}\bar{\eta},\text{ \ }\Omega_{\bar{\eta
}}=\pi_{\bar{\eta}}-\frac{i}{2}\eta,\label{constraints}\\
\Omega_{\chi}  &  =\pi_{\chi}-i\bar{\chi},\text{ \ }\Omega_{\bar{\chi}}%
=\pi_{\bar{\chi}}-i\chi.\nonumber
\end{align}

According to the Dirac formalism, the previous constraints are second class
and the dynamics of the system is obtained when we impose the set of
constraints (\ref{constraints}) and introduce the Dirac brackets, and we
obtain%
\begin{align}
\left\{  a,\pi_{a}\right\}  _{D}  &  =1,\text{ }\left\{  \phi,\pi_{\phi
}\right\}  _{D}=1,\text{ } \left\{  T,\pi_{T}\right\}  _{D}=1,\nonumber\\
\left\{  \lambda,\bar{\lambda}\right\}  _{D}  &  =-\frac{1}{6i},\text{
}\left\{  \chi,\bar{\chi}\right\}  _{D}=-\frac{i}{2},\text{ } \left\{
\eta,\bar{\eta}\right\}  _{D}=-i.
\end{align}

Using the standard definition for the Hamiltonian and imposing the constraints
(\ref{constraints}), we can write the Hamiltonian of the theory as%
\begin{equation}
H=NH_{0}+\frac{1}{2}\psi S-\frac{1}{2}\bar{\psi}\bar{S},\label{susyham}%
\end{equation}
where%
\begin{equation}%
\begin{array}
[c]{c}%
H_{0}=-\frac{\kappa^{2}\pi_{a}^{2}}{12a}+\frac{\kappa^{2}\pi_{T}^{2}}{2a^{3}%
}-\frac{3i\kappa^{2}\pi_{T}}{2a^{3}}\left(  \lambda\bar{\eta}+\bar{\lambda}%
\eta\right)  +\frac{\kappa^{2}\pi_{\phi}^{2}}{4a^{3}}-\frac{3i\kappa^{2}%
\pi_{\phi}}{2a^{3}}\left(  \lambda\bar{\chi}+\bar{\lambda}\chi\right)
-\frac{3a^{3}}{4\kappa^{2}}W^{2}\\
+\frac{3\sqrt{k}a^{2}}{\kappa^{2}}W-\frac{3ka}{\kappa^{2}}+\frac{a^{3}%
}{2\kappa^{2}}W_{T}^{2}+\frac{a^{3}}{4\kappa^{2}}W_{\phi}^{2}+\frac{9}%
{2}W\lambda\bar{\lambda}-\frac{3}{4}W\eta\bar{\eta}-\frac{3}{2}W\chi\bar{\chi
}\\
+\frac{3}{2}W_{T}\left(  \lambda\bar{\eta}-\bar{\lambda}\eta\right)  +\frac
{3}{2}W_{\phi}\left(  \lambda\bar{\chi}-\bar{\lambda}\chi\right)  +W_{TT}%
\eta\bar{\eta}+W_{T\phi}\left(  \chi\bar{\eta}-\bar{\chi}\eta\right)  \\
+W_{\phi\phi}\chi\bar{\chi}-\frac{3\sqrt{k}}{a}\lambda\bar{\lambda}%
+\frac{3\sqrt{k}}{2a}\eta\bar{\eta}+\frac{3\sqrt{k}}{a}\chi\bar{\chi}%
-\frac{9\kappa^{2}}{2a^{3}}\lambda\bar{\lambda}\chi\bar{\chi}-\frac
{9\kappa^{2}}{4a^{3}}\lambda\bar{\lambda}\eta\bar{\eta}-\frac{3\kappa^{2}%
}{4a^{3}}\eta\bar{\eta}\chi\bar{\chi},%
\end{array}
\label{H0_SUSY}%
\end{equation}%
\begin{equation}%
\begin{array}
[c]{c}%
S=\frac{\kappa\pi_{a}}{\sqrt{a}}\lambda+\frac{\kappa\pi_{T}}{\sqrt{a^{3}}}%
\eta+\frac{\kappa\pi_{\phi}}{\sqrt{a^{3}}}\chi-\frac{6i\sqrt{ak}}{\kappa
}\lambda+\frac{3i\sqrt{a^{3}}}{\kappa}W\lambda\\
+\frac{i\sqrt{a^{3}}}{\kappa}W_{T}\eta+\frac{i\sqrt{a^{3}}}{\kappa}W_{\phi
}\chi+\frac{3i\kappa}{2a^{3/2}}\lambda\eta\bar{\eta}+\frac{3i\kappa}{a^{3/2}%
}\lambda\chi\bar{\chi},%
\end{array}
\label{S}%
\end{equation}%
\begin{equation}%
\begin{array}
[c]{c}%
\bar{S}=\frac{\kappa\pi_{a}}{\sqrt{a}}\bar{\lambda}+\frac{\kappa\pi_{T}}%
{\sqrt{a^{3}}}\bar{\eta}+\frac{\kappa\pi_{\phi}}{\sqrt{a^{3}}}\bar{\chi}%
+\frac{6i\sqrt{ak}}{\kappa}\bar{\lambda}-\frac{3i\sqrt{a^{3}}}{\kappa}%
W\bar{\lambda}\\
-\frac{i\sqrt{a^{3}}}{\kappa}W_{T}\bar{\eta}-\frac{i\sqrt{a^{3}}}{\kappa
}W_{\phi}\bar{\chi}-\frac{3i\kappa}{2a^{3/2}}\bar{\lambda}\eta\bar{\eta}%
-\frac{3i\kappa}{a^{3/2}}\bar{\lambda}\chi\bar{\chi},%
\end{array}
\label{bar_S}%
\end{equation}
satisfy the Dirac algebra $\{S,\bar{S}\}_{D}=2 H_{0}$, $\{H_{0},S\}_{D}=\{H_{0},\bar{S}\}_{D}=0$.

\section{Superpotential solutions}

From the $H_{0}$, (\ref{H0_SUSY}), we identify the scalar potential
\begin{equation}
U(a,\phi,T)=-\frac{3k}{a^{2}}+e^{2\phi}V(T),\label{escalar}
\end{equation}
which is related to the superpotential $W(\phi,T) $
by%
\begin{equation}
\label{scalarpotential}U=-\frac{3W^{2}}{4}+\frac{3\sqrt{k}}%
{a}W-\frac{3k}{a^{2}}+\frac{1}{4}\left(  \frac{\partial W}{\partial\phi
}\right)  ^{2}+\frac{1}{2}\left(  \frac{\partial W}{\partial T}\right)  ^{2}.
\end{equation}
The form of the scalar potential (\ref{escalar}) of a FRW geometry is consistent with $k=0$, hence
we restrict ourselves to this geometry, and we get the equation
\begin{equation}
e^{2\phi}V(T)= -\frac{3W^{2}}{4}+\frac{1}{4}\left(  \frac{\partial W}{
\partial\phi}\right)  ^{2}+\frac{1}{2}\left(  \frac{\partial W}{\partial
T}\right)  ^{2}.
\end{equation}
This suggest us a separation of variables of the form
$W(\phi,T)=\frac{1}{\sqrt{2}}e^{\phi}f(T)$. With this ansatz we obtain the
following relation between the tachyon potential and the tachyonic component
of the superpotential
\begin{equation}
\label{eq_pot_spot}(f^{\prime})^{2}-f^{2}=V(T),
\end{equation}
where the prime denotes differentiation with respect to $T$. In order to find solutions to this equation we must fix the
function $V\left(  T\right)  $. For example if $V(T)=0$, the solution is
$f(T)=e^{T}$. Further, for
\begin{equation}
\label{crunch}V(T)=\frac{m^{2}}{2}\left(  -T^{2}+\frac{1}{4}T^{4}\right)  ,
\end{equation}
which according to the analysis of Zwiebach \emph{et.al.}, produces a big
crunch scenario as the final state of the universe \cite{zwiebach}, there is an
imaginary solution $f(T)=imT^{2}/({2\sqrt{2}})$, whose superpotential is
\begin{equation}
W(T)=\frac{i}{4}me^{\phi}T^{2}, \label{solim}%
\end{equation}
which generates complex fermion masses. However, as we show in Appendix
\ref{powerseries}, there is also a real solution given by an
infinite power series, which can be written as (\ref{freal}),
\begin{equation}
W(T)=\frac{1}{\sqrt2}e^{\phi}\left\{  e^{T}-\frac{1}{12}m^{2}T^{3}\left[
1-\frac{1}{2}T+\frac{1}{20}\left(  1+\frac{3}{2}m^{2}\right)  T^{2}%
+\mathcal{O}(T^{3})\right]  \right\}  .
\label{pseries}
\end{equation}

Other proposal are potentials of the form $V(T)=\exp(\nu T)$ \cite{zwiebach}, they are known
to prevent the tachyon from reaching infinity in certain cases, with $\nu
\geq2$ there is no initial (positive) tachyon velocity for which the tachyon
can reach $T = \infty$. For this potential we find
$f(T)=\pm2(\nu^{2}-4)^{-1/2}\exp(\nu T/2)$, and the superpotential is in this
case%
\begin{equation}
\label{spotexp}W(\phi,T)=\pm\frac{2}{\sqrt{\nu^{2}-4}}\exp\left(  \phi
+\frac{\nu}{2}T\right)  , \text{ } \nu\neq2.
\end{equation}

\begin{figure}[H]
\centering
\includegraphics[height=5cm,width=9cm]{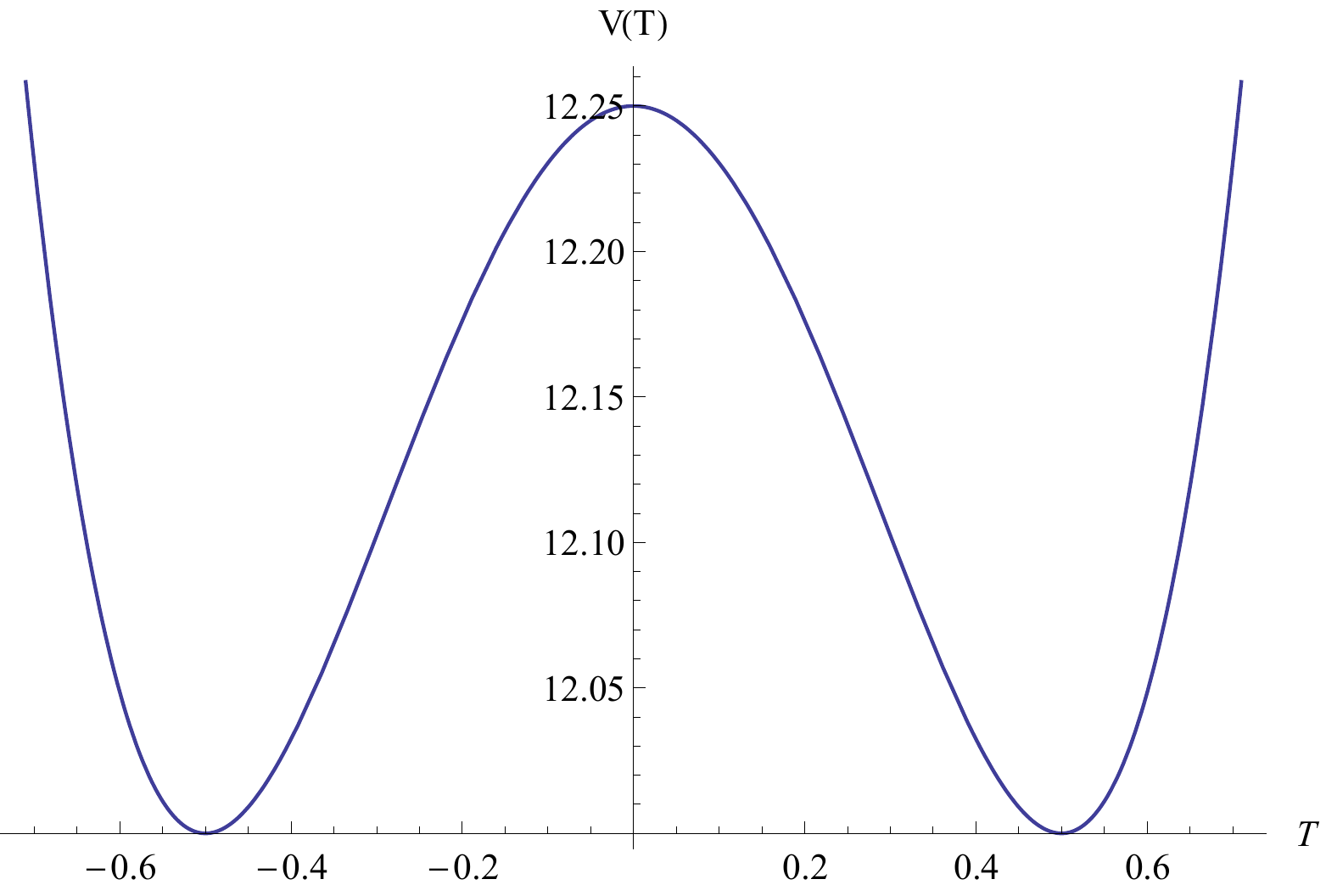}
\caption{{\protect\footnotesize {Tachyon
potential V(T) from eq. (\ref{potentialp4}) for $ \tau = 1/2 $.}}}%
\label{figpotentialp4}%
\end{figure}

Another interesting proposal is for instance
\begin{equation}
\label{spotp4}
W(\phi,T)= e^{\phi}f(T)=\frac{i e^{\phi}[(T-\tau ) (\tau +T)+2]}{2 \tau ^2},
\end{equation}
as in the case of eq. (\ref{solim}) this superpotential generates complex fermion masses, however it can be made to be real in complete analogy with (\ref{pseries}). The tachyon potential corresponding to (\ref{spotp4}) is
\begin{equation}
\label{potentialp4}
V(T)=\frac{1}{4 \tau ^4}\left[\left(\tau ^2-2\right)^2+T^4-2 \tau ^2 T^2\right],
\end{equation}
this potential, shown in Fig. \ref{figpotentialp4}, has a maximum at $ T=0 $, with $ V(0)= 1/4 -1/\tau^{2} + 1/\tau^{4} $, and minima at $ T= \pm \tau $, with $ V(\tau)=V(-\tau)= -1/\tau^{2} + 1/\tau^{4} $, it also holds that $ V(0)- V(\pm \tau)= 1/4 $, thus if we let $ \tau \to \infty $ the potential difference will remain the same, as in the case of Sen's conjectures.
If we compute the superpotential corresponding to (\ref{potentialp4}) following Appendix B, it can be shown that $W_T(\phi,0)=e^\phi\sqrt{4(1-\tau^2)+ \tau^4(A^2-1)}/\tau^2$, where $A=f(0)$. Further, at the minimum, i.e. at $T=\tau$, we make the power expansion around this point and we get $W_T(\phi,\tau)=e^\phi \sqrt{4(1-\tau^2)+ \tau^4(B^2-2)}/\tau^2$, where $B=f(\tau)$. If we take the limit $\tau\to\infty$, in order to make the computation we set $u=1/T$ and $\vartheta=1/\tau$ and evaluate the result first at $u=0$ and then we set $\vartheta=0$, we get $W_T(\phi,u)|_{u=0,\vartheta=0}=e^\phi\sqrt{C^2-1}$, where $C=f(u)_{u=0,\vartheta=0}$. Thus if we set $C=1$, the minimum in this limit is supersymmetric. In the limit $\tau\to\infty$ we have also $W_T(\phi,T)|_{T=0}=e^\phi\sqrt{A^2-1}$

hence it does not vanish and supersymmetry is broken. Further, for $T=\tau$ we can make the power series ansatz in the neighborhood of this value, and we get $W_T(\phi,\tau)=b$, where $b$ is another constant and it can be easily seen that if $b=0$, the whole power series tends to zero when $\tau\to\infty$, i.e. in this limit $W_T(\phi,T)=0$ and supersymmetry is conserved.

An interesting question regards potentials with suitable supersymmetric
properties, of the type of Sen conjectures. For instance the potential
\begin{equation}
\label{potexppoly4_alpha}V(T)=\exp(-nT)\left[  \alpha_{0}+\alpha_{1}%
T+\alpha_{2}T^{2}+\alpha_{3}T^{3}+\alpha_{4}T^{4}\right]  ,
\end{equation}
in this case we have $f(T)=(a+b T+c T^{2})\exp\left(  - nT/2\right)  $ from which we obtain 
\begin{equation}
\label{spotexppoly4abc}
W(T,\phi)=(a+b T+c T^{2})\exp\left(  \phi- nT/2\right),
\end{equation}
the explicit coefficients $\alpha_{i}$ depend on the free parameters $a$, $n$ and $V_{0}$, in fact we demand the presence of a maximum for $T=0$, this provides us with two equations which can be solved for $b$ and $c$ in (\ref{spotexppoly4abc}) and with the condition $V(0)<0$, details of the calculations are given in Appendix (\ref{apexppoly4}), see Fig. \ref{figpotential}.
\begin{figure}[H]
\centering
\includegraphics[height=5cm,width=9cm]{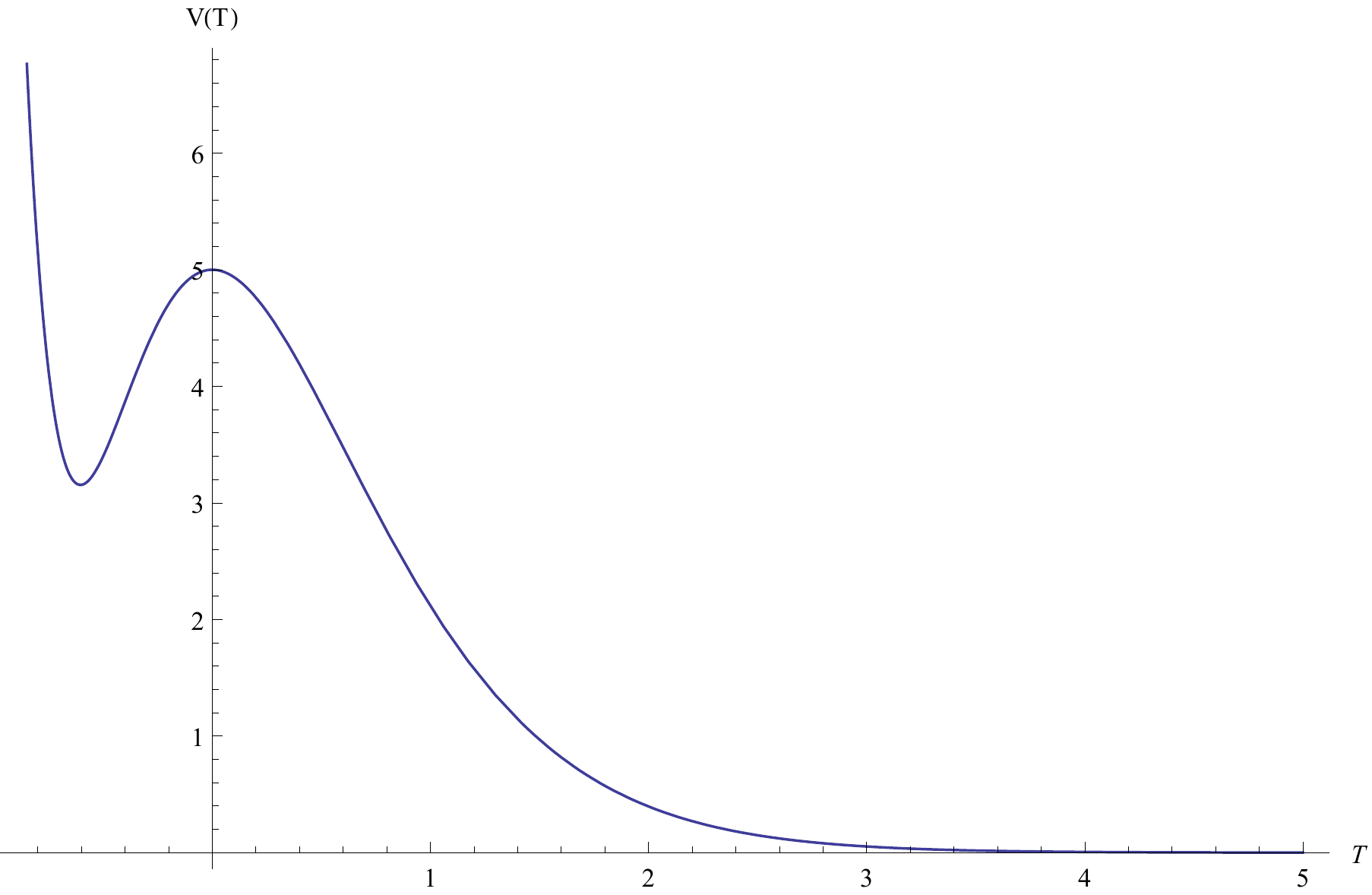}
\caption{{\protect\footnotesize {Tachyon
potential V(T) produced by $f(T)$ in (\ref{fexppoly4}) for $a=10$, $n=3$ and $V_{0}=5$.}}}%
\label{figpotential}%
\end{figure}

At the maximum of this potential $W_{T}(0,\phi) \neq0$ and $W_{\phi}(0,\phi) \neq0$,
hence supersymmetry is broken. Further,  after condensation supersymmetry is restored
because $W_{T}\to0$ and $W_{\phi} \to0$, when $T \to\infty$.


\section{Conclusions}

We have studied a worldline supersymmetric theory in a FRW background,
with a closed string tachyon. We have constructed the action in the formalism of the
`new'-$\Theta$ variables in one dimension, which allows systematically to construct
supergravity actions. We consider the solutions for the differential equation of the superpotential for given tachyonic
potentials and we have obtained solutions with broken supersymmetry at the unstable,
tachyonic, configuration, and supersymmetric at the stable minimum. Furthermore, the
superpotentials can have simple forms, but which correspond to complex fermionic masses.
These superpotentials can be obtained as well by a power series ansatz, whose general solution depends on a real parameter which can be chosen such that the complex solutions can be mapped to real solutions. Some of these potentials have been considered in cosmological models, like in \cite{zwiebach} where inflationary and big crunch sceneries are given, and it would be interesting to consider the supersymmetric versions.

\vskip 1truecm
\centerline{\bf Acknowledgments} We thank BUAP-VIEP and PIFI-SEP for the
support, V. V\'{a}zquez-B\'{a}ez thanks CONACyT for the studies grant during
this work.


\appendix

\section{Superspace supergravity}

\label{apsugra} In this appendix we shortly review the `new' superspace
formulation of supergravity following \cite{cupa}. Superfields are defined as
\begin{equation}
\phi(z)\rightarrow\Phi(z)=e^{\Theta^{\alpha}\mathcal{D} _{\alpha}} \phi(z)
\vert_{\theta=0}, \label{new}%
\end{equation}
where $\Theta$ are anticommuting Lorentz ($SL(2,C)$) covariant spinorial
variables. In order to ensure full covariance for these new superfields, the
hole ``old" superspace can be kept, setting $\theta=0$ at the end of the
computations, i.e. $\Phi(z,\Theta)= e^{\Theta^{\alpha}\mathcal{D} _{\alpha}}
\phi(z)$.

The preceding redefinition of superfields is complemented by the usual
redefinition of local supersymmetry transformations in such a way that Lorentz
covariance is kept. The way is to add a local Lorentz transformation to the
local superspace translations as follows\cite{wztrafo}:
\begin{equation}
\delta_{\xi}\phi_{A}(z)= -\xi^{B} {E_{B}}^{M} \left(  \partial_{M} \phi_{A}-
{\phi_{M A}}^{B} \phi_{B} \right)  =- \xi^{B}\mathcal{D}_{B} \phi_{A}(z),
\end{equation}
hence the new superfields, whose components are Lorentz covariant, transform
as $\delta_{\xi}\Phi(z, \Theta)=-\xi^{A} \mathcal{D}_{A}\Phi(z, \Theta)$,
i.e.
\begin{equation}
\delta_{\xi}\Phi_{A}(z, \Theta)= -\xi^{B} \mathcal{D}_{B}\phi_{A}-
\Theta^{\beta}\xi^{B} \mathcal{D}_{B} \mathcal{D}_{\beta}\phi_{A} -\frac{1}{2}
\Theta^{\beta_{1}} \Theta^{\beta_{2}} \xi^{B} \mathcal{D}_{B} \mathcal{D}%
_{\beta_{1}} \mathcal{D}_{\beta_{2}} \phi_{A}+ \dots\label{trafo}%
\end{equation}
The computation of this expression is done taking into account the fact that
the multiple covariant derivatives arising from the exponential in (\ref{new})
appear as fully antisymmetrized products. Thus, when a further derivative is
applied on this product, the result must be antisymmetrized, e.g.
\begin{equation}
\mathcal{D}_{\alpha}\mathcal{D}_{\beta}\phi_{A}={\frac{1}{2}} \{\mathcal{D}%
_{\alpha}, \mathcal{D}_{\beta}\} \phi_{A} + {\frac{1}{2}} [\mathcal{D}%
_{\alpha}, \mathcal{D}_{\beta}] \phi_{A}=-{T_{\alpha\beta}}^{C} \mathcal{D}%
_{C} \phi_{A}-{R_{\alpha\beta A}}^{B}\phi_{B}+ \mathcal{D}_{[\alpha
}\mathcal{D}_{\beta]} \phi_{A}, \label{dd}%
\end{equation}
where the last term is precisely the second order term of $\Phi(z, \Theta)$.
Following these lines, it can be shown that (\ref{trafo}) can be cast into the
form
\begin{equation}
\delta_{\xi}\Phi_{A}(z, \Theta)= {\eta_{\xi}}^{\alpha}(z, \Theta)
{\frac{\partial}{{\partial\Theta^{\alpha}}}}\Phi_{A}(z, \Theta)+ {\eta_{\xi}%
}^{a}(z, \Theta) \mathcal{D}_{a}\Phi_{A}(z, \Theta)+ {\eta_{\xi A}}^{B}(z,
\Theta)\Phi_{B}(z, \Theta), \label{geom}%
\end{equation}
where the coefficients ${\eta_{\xi}}^{A}(z, \Theta)$ and ${\eta_{\xi A}}%
^{B}(z, \Theta)$ depend on components of the curvature and torsion tensors and
their covariant derivatives.

In order to have a geometric formulation in the new superspace, following the
Wess-Zumino gauge \cite{wzgauge}, which eliminates the gauge degrees of
freedom introduced by the generalization of local supersymmetry to superspace
diffeomorphisms, a new vielbein is introduced. Let us consider a vector field
$V_{m}={E_{m}}^{a}V_{a}+{E_{m}}^{\alpha}V_{\alpha}$, this relation can be
inverted to
\begin{equation}
V_{a}=E_{\quad a}^{(-1)m}(V_{m}-{E_{m}}^{\alpha}V_{\alpha})=\tilde
E_{a}^{\ \tilde M}V_{\tilde M},%
\end{equation}
where the indices $\tilde{M} \equiv(m, \alpha)$ contain a spacetime world
index and a spinorial local index, i.e. $\tilde E_{a}^{\ \tilde M}=(\tilde
E_{a}^{\ m},\tilde E_{a}^{\ \alpha})\equiv(E_{\quad a}^{(-1)m},-E_{\quad
a}^{(-1)m}{E_{m}}^{\alpha})$. With this definition and $\tilde E_{\alpha
}^{\ \tilde M}=\delta_{\alpha}^{\ \tilde M}$, an inverse vielbein $\tilde
E_{A}^{\ \tilde M}$ can be defined. The corresponding vielbein is then
\begin{equation}
{\tilde{E}_{\tilde M}}^{\ \ B}=\left(
\begin{array}
[c]{ll}%
{E_{m}}^{\, b} & {E_{m}}^{\, \beta} \nonumber\\
0 & {\delta_{\alpha}}^{\, \beta} \nonumber
\end{array}
\right),
\end{equation}
i.e. $\tilde E_{A}^{\ \tilde M}{\tilde{E}_{\tilde M}}^{\ \ B}=\delta_{A}%
^{\ B}$ and ${\tilde{E}_{\tilde M}}^{\ \ A}\tilde E_{A}^{\ \tilde N}%
=\delta_{\tilde M}^{\ \tilde N}$. Even if this vielbein seems to correspond to
the Wess-Zumino gauge, full covariance can be kept by considering certain
components of the torsion and curvature as independent degrees of freedom of
the supergravity multiplet. If we define in this basis covariant derivatives
as usual by $\mathcal{D}_{A}V_{\tilde M}=(-1)^{(m+b)a}{\tilde{E}_{\tilde M}%
}^{\ \ B}\mathcal{D}_{A}V_{B}$ and $\mathcal{D}_{A}V^{\tilde M}=\mathcal{D}%
_{A}V^{B}{\tilde{E}_{B}^{\ \tilde M}}$, then
\begin{equation}
\label{conm}[\mathcal{D}_{\tilde M},\mathcal{D}_{\tilde N}]_{\pm}%
V_{A}=-{T_{\tilde M\tilde N}}^{\tilde P}\mathcal{D}_{\tilde P}V_{A}-{R_{\tilde
M\tilde N A}}^{B}V_{B},
\end{equation}
which supplemented by the corresponding Bianchi identities, contains all the
information of supergravity. As (\ref{conm}) does not contain derivatives of
the old $\theta$-variables, the different levels in the $\theta$-expansion
decouple, and the limit $\theta^{\mu}=0$ does not require gauge fixing.
Actually, supergravity transformations can be written as $\delta_{\xi}\Phi(z,
\Theta)= -\tilde\xi^{\tilde M} \tilde{\mathcal{D}}_{\tilde M}\Phi(z, \Theta)$
and following the same lines as for (\ref{geom}) we get
\begin{equation}
\delta_{\xi}\Phi(z, \Theta)= \left[  {\tilde\eta_{\xi}}^{\ \alpha}(z, \Theta)
{\frac{\partial}{{\partial\Theta^{\alpha}}}}+ {\tilde\eta_{\xi}}^{\ m}(z,
\Theta) \mathcal{D}_{m}\right]  \Phi(z, \Theta). \label{geom1}%
\end{equation}
Further, the covariant derivative $\mathcal{D}_{m}$ on the r.h.s. of this
expression acts on the components of the superfield $\Phi(z,\Theta)$ as in
(\ref{trafo}), which can be written as
\begin{align}
\mathcal{D}_{m}\Phi(z, \Theta)  &  = {\partial}_{m}\phi+ \Theta^{\beta
}({\partial}_{m}-\omega_{m\beta}^{\ \ \ \gamma} )\mathcal{D}_{\gamma}%
\phi-\frac{2}{2} \Theta^{\beta_{1}} \Theta^{\beta_{2}} ({\partial}_{m}%
-\omega_{m\beta_{1}}^{\ \ \ \gamma} )\mathcal{D}_{\gamma} \mathcal{D}%
_{\beta_{2}} \phi+ \dots\nonumber\\
&  ={\partial}_{m}\Phi(z, \Theta)+\Theta^{\beta}\omega_{m\beta}^{\ \ \ \gamma
}\partial_{\gamma}\Phi(z, \Theta). \label{dm}%
\end{align}
Therefore, including a Lorentz index, (\ref{geom1}) can be written as
\begin{equation}
\delta_{\xi}\Phi_{A}(z, \Theta)={\hat\eta_{\xi}}^{\ \tilde{M}}(z, \Theta)
\partial_{\tilde{M}} \Phi_{A}(z, \Theta)+{\hat\eta_{\xi A}}^{\ \ B}(z, \Theta)
\Phi_{B}(z, \Theta), \label{trafos}%
\end{equation}
where $\hat\eta_{\xi}^{\ \alpha}=\tilde\eta_{\xi}^{\ \alpha}-\tilde\eta_{\xi
}^{\ m}\Theta^{\beta}\phi_{m\beta}^{\ \ \ \alpha}$, ${\hat\eta_{\xi}}%
^{\ m}={\tilde\eta_{\xi}}^{\ m}$ and ${\hat\eta_{\xi A}}^{\ \ B}$ is Lie
algebra valued. Further, ${\tilde\eta_{\xi}}^{\tilde M}=\tilde\xi^{\tilde
N}{\tilde\eta_{\tilde N}}^{\tilde M}$ and ${\tilde\eta_{m}}^{\tilde N}%
={\delta_{m}}^{\tilde N}$, ${\tilde\eta_{mD}}^{\ \ \ B}=0$ and ${\tilde\eta
}_{\alpha}^{\prime{\tilde N}}={\tilde\eta}_{\alpha}^{\tilde N}-\delta_{\alpha
}^{\tilde N}$ can be obtained from the following recursion relation
\begin{align}
\left(  1+\Theta^{\beta}\frac{\partial}{\partial\Theta^{\beta}}\right)
{\tilde\eta}_{\alpha}^{\prime{\tilde N}}=\Theta^{\beta}\mathcal{D}_{\beta
}{\tilde\eta}_{\alpha}^{\tilde N}+\Theta^{\alpha_{1}}\left(  -\Theta
^{\alpha_{2}}{R_{\alpha_{2}\alpha\alpha_{1}}}^{\tilde N}+{T_{\alpha\alpha_{1}%
}}^{\tilde L}{\tilde\eta}_{\tilde L}^{\tilde N}\right) \nonumber\\
+{\tilde\eta}_{\alpha}^{m}\Theta^{\alpha_{1}}\left(  -\Theta^{\alpha_{2}%
}{R_{\alpha_{2}m\alpha_{1}}}^{\tilde N}+{T_{m\alpha_{1}}}^{\tilde L}%
{\tilde\eta}_{\tilde L}^{\tilde N}\right)  -{{\tilde\eta}_{\alpha}}%
^{\prime\gamma}{{\tilde\eta}_{\gamma}}^{\prime{\tilde N}}, \label{recursion}%
\end{align}
and a similar one for ${\tilde\eta_{\xi A}}^{\ \ B}$. It turns out that
\begin{align}
{\tilde\eta_{\xi}}^{\ \tilde{M}}(z, \Theta)  &  =-\tilde\xi^{\tilde M}%
+\Theta^{\gamma}\left(  \frac{1}{2}\tilde\xi^{\beta}{T_{\gamma\beta}}^{\tilde
M}+\tilde\xi^{n}{\phi_{n\gamma}}^{\tilde M}\right) \nonumber\\
&  -\frac{1}{2}\Theta^{\gamma}\Theta^{\delta}\left(  -\frac{2}{3}\tilde
\xi^{\beta}{R_{\delta\beta\gamma}}^{\tilde M}+\frac{1}{3}\mathcal{D}_{\delta
}{T_{\gamma\beta}}^{\tilde M}+ {T_{\delta\beta}}^{\tilde n}{\phi_{n\gamma}%
}^{\tilde M}-\frac{1}{3}{T_{\delta\beta}}^{\tilde n}{T_{n\gamma}}^{\tilde
M}+\frac{1}{6}{T_{\delta\beta}}^{\tilde\epsilon}{T_{\epsilon\gamma}}^{\tilde
M}\right)  +\cdots\label{eta}%
\end{align}
Consistently with these ideas, covariant derivatives can be defined as
\begin{equation}
\nabla_{A}= e^{\Theta^{\alpha}\mathcal{D}_{\alpha}} \mathcal{D}_{A}%
e^{-\Theta^{\alpha}\mathcal{D}_{\alpha}}.
\end{equation}
Following a similar reasoning as the one which lead to (\ref{trafos}), it can
be shown that
\begin{equation}
\nabla_{A}\Phi_{B}={\nabla_{A}}^{\tilde M} \partial_{\tilde M} \Phi
_{B}+{\nabla_{AB}}^{C} \Phi_{C}, \label{covder}%
\end{equation}
where ${\nabla_{A}}^{\tilde M}$ is the inverse vielbein of the new superspace
and if we write it as ${\nabla_{A}}^{\tilde M}={\tilde E_{A}}^{\tilde
N}{\nabla_{\tilde N}^{\ \tilde M}}$, it can be obtained from the recursion
relations
\begin{equation}
\left(  {\delta_{\tilde M}}^{\gamma}{\delta_{\gamma}}^{\tilde L}%
+{\delta_{\tilde M}}^{\tilde L}\Theta^{\beta}\frac{\partial}{\partial
\Theta^{\beta}}\right)  {{\nabla^{\prime}}_{\tilde L}}^{\tilde N}=-{T_{\tilde
M}}^{\tilde L}{\nabla_{\tilde L}}^{\tilde N}-{{\nabla^{\prime}}_{\tilde M}%
}^{\gamma}{{\nabla^{\prime}}_{\gamma}}^{\tilde N}-(-1)^{m}\Theta^{\alpha_{1}%
}{\nabla_{\tilde M\alpha_{1}}}^{\gamma}{\nabla_{\gamma}}^{\tilde N},
\end{equation}
and a similar one for ${\nabla_{AB}}^{C}$. The vielbein ${\nabla_{\tilde M}%
}^{A}$, i.e. ${\nabla_{\tilde M}}^{A}{\nabla_{A}}^{\tilde N}=\delta_{\tilde
M}^{\tilde N}$, transforms as
\begin{equation}
\delta_{\xi}{\nabla_{\tilde M}}^{A}=\partial_{\tilde M}\hat\eta_{\xi}^{\tilde
N}{\nabla_{\tilde N}}^{A}+\hat\eta_{\xi}^{\tilde N}\partial_{\tilde N}%
{\nabla_{\tilde M}}^{A} -{\nabla_{\tilde M}}^{B}{\hat\eta_{\xi B}}^{\ \ A},
\label{trafov}%
\end{equation}
and to second order is given by
\begin{align}
{\nabla_{m}}^{B}  &  ={E_{m}}^{B}+\Theta^{\gamma}({T_{\gamma m}}^{B}+{\phi_{m
\gamma}}^{B})+\frac{1}{2}\Theta^{\gamma}\Theta^{\delta}\left(  -{R_{m\delta
\gamma}}^{B}+\mathcal{D}_{\delta}{T_{\gamma m}}^{B}+{T_{m\delta}}%
^{A}{T_{A\gamma}}^{B}-{\phi_{m\delta}}^{\beta}{T_{\gamma\beta}}^{B}\right)  +
\cdots\nonumber\\
{\nabla_{\alpha}}^{B}  &  ={\delta_{\alpha}}^{B}+\frac{1}{2}\Theta^{\gamma
}{T_{\gamma\alpha}}^{B} +\frac{1}{6}\Theta^{\gamma}\Theta^{\delta}\left(
-{R_{\delta\alpha\gamma}}^{B}+2\mathcal{D}_{\delta}{T_{\gamma\alpha}}%
^{B}+{T_{\delta\alpha}}^{D}{T_{D\gamma}}^{B}\right)  +\cdots\label{nabla}%
\end{align}
As in ordinary supergravity, the superdeterminant of the vielbein is an
invariant density
\begin{equation}
\mathcal{E}=\mathrm{Sdet}\left(  {\nabla_{\tilde M}}^{A}\right)
\equiv{{det\left(  {\nabla_{m}}^{a} - {\nabla_{m}}^{\beta}{{\nabla^{(-1)}%
}_{\beta}}^{\gamma}{\nabla_{\gamma}}^{a}\right)  }/{det\left(  {\nabla
_{\alpha}}^{\beta}\right)  }},%
\end{equation}
which transforms as
\begin{equation}
\delta_{\xi}\mathcal{E}= (-1)^{m}\partial_{\tilde M} \left(  \eta_{\xi
}^{\tilde M} \mathcal{E}\right),  \label{trafoe}%
\end{equation}
and the superspace integral of the product of the invariant density with any
Lorentz invariant superfield will be by construction invariant under
supergravity transformations.

Therefore, local supersymmetry can be formulated in the new superspace in a
geometrical way, with the only difference that now the transformation
parameters are field dependent, depending on components of the torsion and
curvature and their covariant derivatives, subject to the Bianchi identities.
This formulation is manifestly covariant in the framework of the highly
redundant superspace $(z,\Theta)$. However, as in the transformations
(\ref{conm}), (\ref{trafos}), (\ref{trafov}) and (\ref{trafoe}) there are no
derivatives of the old $\theta^{\mu}$-variables, they can be set to zero
without loss of generality.

\section{Power series ansatz}

\label{powerseries} Equation (\ref{eq_pot_spot}) can be solved by power series
ansatz. Let us set $V(T)=\sum_{l\geq0}v_{l} T^{l}$ and $f(T)=\sum_{l\geq
0}f_{l} T^{l}$, then
\begin{equation}
V(T)=\sum_{l\geq0}\left[  \sum_{m=0}^{l+2}m(l-m+2)f_{m}f_{l-m+2}T^{l}%
-\sum_{m=0}^{l} f_{m}f_{l-m}\right]  T^{l},
\end{equation}
that is
\begin{equation}
v_{l}=2(l+1)f_{1}f_{l+1}+\sum_{m=2}^{l} m(l-m+2)f_{m}f_{l-m+2}-\sum_{m=0}^{l}
f_{m}f_{l-m},
\end{equation}
which can be solved as follows. If $f_{1}=\pm\sqrt{v_{0}+f_{0}^{2}}$ does not
vanish, then for $l>1$, $f_{l+1}$ can be obtained in terms of $v_{l}$ and
$f_{l}$,
\begin{equation}%
\begin{array}
[c]{ll}%
f_{2} & =\frac{1}{4f_{1}}(2f_{0}f_{1}+v_{1}),\\
f_{3} & =\frac{1}{6f_{1}}(f_{1}^{2}+2f_{0}f_{2}-4f_{2}^{2}+v_{2}),\\
f_{4} & =\frac{1}{6f_{1}}(2f_{1}f_{2}+2f_{0}f_{3}-12f_{2}f_{3}+v_{3}).\\
& \vdots
\end{array}
\label{sol1}%
\end{equation}
This solution depends on the free parameter $f_{0}$ and in general is singular
in $f_{1}$. For example, in the case of the exponential potential,
$V(T)=e^{2\kappa T}$, it can be verified that (\ref{sol1}) coincides with
$f(T)=\frac{1}{\sqrt{\kappa^{2}-1}}e^{\kappa T}$, with $f_{0}^{2}%
=1/(\kappa^{2}-1)$. Further, in the singular case when $f_{1}=0$, which
corresponds to $f_{0}=\pm\sqrt{-v_{0}}$, we see from the first equation of
(\ref{sol1}), that there are solutions only if $v_{1}=0$. In this case we get
\begin{equation}%
\begin{array}
[c]{ll}%
f_{2} & =\frac{1}{4}(f_{0}\pm\sqrt{f_{0}^{2}+4v_{2}}),\\
f_{3} & =-\frac{v_{3}}{2(f_{0}-6f_{2})},\\
f_{4} & =\frac{1}{2(f_{0}-8f_{2})}(-f_{2}^{2}+9f_{3}^{2}-v_{4}).\\
& \vdots
\end{array}
\label{sol2}%
\end{equation}
The square roots in these solutions can lead to imaginary terms, similarly to
the `imaginary mass' of the tachyon. Such problems can be avoided if the
integration constant $f_{0}$ is suitably chosen, as can be seen for the
potential $V(T)=v_{2}T^{2}+v_{4}T^{4}$. In this case $f_{1}=f_{0}$ and
$f_{0}(2f_{2}-f_{0})=0$, and if we choose $f_{0}=0$, then $f_{1}=0$,
$f_{2}=\pm\frac{1}{2}\sqrt{v_{2}}$, $f_{3}=\frac{1}{12f_{2}}v_{3}$,
$f_{4}=\frac{1}{16f_{2}}(f_{2}^{2}-9f_{3}^{2}+v_{4})$, $f_{5}=\frac{1}%
{20f_{2}}(f_{2}f_{3}-24f_{3}f_{4}+v_{5})$, etc. This is the situation of
(\ref{crunch}), where $v_{2}<0$ and $v_{4}=-v_{2}/4$, hence $f_{2} $ becomes
imaginary and $f_{l}=0$ for $l>3$, as in (\ref{solim}). However, if we keep
$f_{0}\neq0$, then $f_{1}=f_{0} $, and from equations (\ref{sol1}) we get
another solution which, setting $v_{2}=-m^{2}/2$, is given by an infinite
series
\begin{equation}
\label{freal}f(T)=f_{0}\bigg[1+T+\frac{1}{2}T^{2}+\frac{1}{3!}\left(
1-\frac{m^{2}}{2f_{0}^{2}}\right)  T^{3} +\frac{1}{4!}\left(  1+\frac{m^{2}%
}{f_{0}^{2}}\right)  T^{4}+\frac{1}{5!}\left(  1-\frac{m^{2}}{2f_{0}^{2}%
}-\frac{3m^{4}}{4f_{0}^{4}}\right)  T^{5}+\cdots\bigg].
\end{equation}


\section{Potential (\ref{potexppoly4_alpha})}

\label{apexppoly4}

We are interested on potentials fulfilling the type of 
requirements of Sen's conjectures. We start from a superpotential of the form 
$f(T)=\exp{[-nT/2}(a+b T+c T^{2})]$, i.e. it rolls down to zero when $T\to\infty$. 
By means of (\ref{eq_pot_spot}) we compute 
the corresponding scalar potential which has the form $V(T)=\exp(-nT)\left[  \alpha_{0}+\alpha
_{1}T+\alpha_{2}T^{2}+\alpha_{3}T^{3}+\alpha_{4}T^{4}\right]  $. Imposing the conditions that 
$V(0)=V_{0}>0$ and $V^{\prime}(0)=0$ we get
\begin{equation}
\label{fexppoly4}f_\pm(T)= e^{-\frac{1}{2} (n x)} \left[  a+\frac{1}{2} \left(  a
n \pm2 \sqrt{a^{2}+V_{0}}\right)  x + \frac{a n^{2} \sqrt{a^{2}+V_{0}}+4 a
\sqrt{a^{2}+V_{0}} \pm4n( a^{2} + V_{0} )}{8 \sqrt{a^{2}+V_{0}}}x^{2}\right].
\end{equation}
If we choose $f_-(T)$ (it would be similar for $f_+$), we get for the parameters of $V(T)$
\begin{equation}%
\begin{array}
[c]{ll}%
\alpha_{0} & =V_{0},\\
\alpha_{1} & =nV_{0},\\
\alpha_{2} & =\left\{
\begin{array}
[c]{l}%
\frac{n^{2}V_{0}^{2}}{a^{2}+V_{0}}+\frac{1}{8}an^{3}\sqrt{a^{2}+V_{0}}%
+\frac{2a^{2}n^{2}V_{0}}{a^{2}+V_{0}}+\frac{5}{2}an\sqrt{a^{2}+V_{0}}\\
-\frac{anV_{0}}{\sqrt{a^{2}+V_{0}}}-\frac{7a^{2}n^{2}}{4}-a^{2}+\frac
{a^{4}n^{2}}{a^{2}+V_{0}}-\frac{a^{3}n}{\sqrt{a^{2}+V_{0}}}-\frac{5n^{2}V_{0}%
}{4}-V_{0},
\end{array}
\right. \\
\alpha_{3} & =\left\{
\begin{array}
[c]{l}%
-nV_{0}+\frac{n^{3}V_{0}}{4}-\frac{n^{3}V_{0}^{2}}{2\left(  a^{2}%
+V_{0}\right)  }-\frac{1}{16}an^{4}\sqrt{a^{2}+V_{0}}+\frac{an^{4}V_{0}%
}{8\sqrt{a^{2}+V_{0}}}-\frac{a^{2}n^{3}V_{0}}{a^{2}+V_{0}}\\
+\frac{3an^{2}V_{0}}{2\sqrt{a^{2}+V_{0}}}+a\sqrt{a^{2}+V_{0}}-2a^{2}%
n-\frac{a^{4}n^{3}}{2\left(  a^{2}+V_{0}\right)  }+\frac{a^{3}n^{4}}%
{8\sqrt{a^{2}+V_{0}}}+\frac{3a^{3}n^{2}}{2\sqrt{a^{2}+V_{0}}},
\end{array}
\right. \\
\alpha_{4} & =\left\{
\begin{array}
[c]{l}%
\frac{n^{4}V_{0}^{2}}{16\left(  a^{2}+V_{0}\right)  }-\frac{n^{2}V_{0}^{2}%
}{4\left(  a^{2}+V_{0}\right)  }-\frac{an^{5}V_{0}}{32\sqrt{a^{2}+V_{0}}%
}+\frac{a^{2}n^{4}V_{0}}{8\left(  a^{2}+V_{0}\right)  }-\frac{a^{2}n^{2}V_{0}%
}{2\left(  a^{2}+V_{0}\right)  }+\frac{anV_{0}}{2\sqrt{a^{2}+V_{0}}}%
+\frac{a^{2}n^{6}}{256}\\
+\frac{a^{2}n^{4}}{64}-\frac{a^{2}n^{2}}{16}-\frac{a^{2}}{4}+\frac{a^{4}n^{4}%
}{16\left(  a^{2}+V_{0}\right)  }-\frac{a^{4}n^{2}}{4\left(  a^{2}%
+V_{0}\right)  }-\frac{a^{3}n^{5}}{32\sqrt{a^{2}+V_{0}}}+\frac{a^{3}n}%
{2\sqrt{a^{2}+V_{0}}}.
\end{array}
\right.
\end{array}
\label{alpha}%
\end{equation}

Now we look for a potential of the form of Fig. \ref{figpotential}, so we must have $n>0$. We require also that
$V^{\prime\prime}(0)<0$ and in addition, for
convenience, we set $\alpha_{i}>0$ resulting in the following constraints for
the parameters $a$, $n$ and $V_{0}$,
\[
n>2, \text{ }a>0, \text{ } \frac{a^{2} n^{6}-12 a^{2} n^{4}+48 a^{2} n^{2}-64
a^{2}}{36 n^{4}+96 n^{2}+64}<V_0<\frac{a^{2} n^{4}-8 a^{2} n^{2}+16 a^{2}}{16
n^{2}}.
\]
Within this rank are located the potentials with profiles like the one
in Fig. \ref{figpotential}.

\vskip 2truecm


\end{document}